\documentclass[preprintnumbers,amsmath,amssymb,showpacs]{revtex4}
\usepackage{graphicx}
\usepackage{dcolumn}
\usepackage{bm}
\usepackage{booktabs}
\usepackage{CJK}
\usepackage{subfigure}
\usepackage{qcircuit}
\usepackage{color}
\usepackage{braket}
\usepackage{listings}
\usepackage{soul}
\usepackage{xcolor}
\soulregister\cite7 
\soulregister\ref7 
\usepackage{color, soul} 
\setstcolor{yellow}

\begin{document}

\title{$(n,m,p)$-type quantum network configuration and its nonlocality}

\author{Zan-Jia Li$^{1}$}
\author{Ying-Qiu He$^{2}$}
\author{Dong Ding$^{2}$}
\email{dingdong@ncist.edu.cn}
\author{Ming-Xing Yu$^{2}$}
\author{Ting Gao$^{3}$}
\email{gaoting@hebtu.edu.cn}
\author{Feng-Li Yan$^{1}$}
\email{flyan@hebtu.edu.cn}

\affiliation {
$^1$ College of Physics, Hebei Normal University, Shijiazhuang 050024, China\\
$^2$ College of Science, North China Institute of Science and Technology, Beijing 101601, China\\
$^3$ School of Mathematical Sciences, Hebei Normal University, Shijiazhuang 050024, China}

\date{\today}

\begin{abstract}

A quantum network shared entangled sources among distant nodes enables us to distribute entanglement along the network by suitable measurements.
Network nonlocality means that it does not admit a network model involving local variables emitted from independent sources.
In this work, we construct an $(n,m,p)$-type quantum network configuration and then derive the corresponding $n$-local correlation inequalities based on the assumption of independent sources.
As a universal acyclic network configuration, it can cover most of the existing network models, such as the typical chain-network and star-network, and admit both centerless and asymmetric configurations.
Then we demonstrate the non-$n$-locality of the present network by calculating the violation of the $n$-local inequality with bipartite entangled sources and Pauli measurements.

\end{abstract}

\pacs{03.65.Ud; 03.67.-a; 03.67.Hk}


%
\maketitle

\section{Introduction}
Bell's theorem \cite{Physics.1(3):195-200(1964)} is one of the most important developments in quantum theory.
The nonlocality of quantum theory \cite{Rev.Mod.Phys.86:419(2014),Rev.Mod.Phys.82.665(2010),Rev.Mod.Phys.81:865(2009)} is the core of the Bell's theorem and it plays an important role in quantum computation and communication \cite{NC2000,Phys.Rev.Lett.87.117901(2001),Phys.Rev.Lett.97.120405(2006),M2015}. Quantum nonlocality is essentially a correlation between the distant parties assumed to be measured by choosing various measurement settings \cite{GYE-PRL2014,DHYG-JPA2020}.
Once the correlation inequalities associated with local theory are violated by any quantum behaviors, the nonlocality of the quantum system will be revealed automatically.

In 2010, Branciard \emph{et al} \cite{Phys.Rev.Lett.104:170401(2010)} proposed a scenario to characterize non-bilocal correlation via entanglement swapping \cite{Phys.Rev.Lett.71.4287(1993)}.
And then, the concept of Bell nonlocality has been generalized to the network nonlocality \cite{PhysRevA.85.032119(2012),PhysRevLett.120.140402(2018),Tavakoli-network2022,PhysRevLett.128.010403(2022),Phys.Rev.A.105:042436.(2022)}, where separated sources are assumed to be independent.
Subsequently, the nonlocality of various quantum networks has been studied, such as $n$-local chain-network configuration \cite{Quantum Inf.Process.14.2025(2015),PhysRevA.102.052222,PhysRevA.107.032404 (2023),PhysRevA.108.032416 (2023)}, star-network configuration \cite{PhysRevA.90.062109(2014),New J.Phys.19.073003(2017),New J.Phys.19.113020.(2017),PhysRevA.108.042409 (2023)}, triangle-network configuration \cite{New J.Phys.14.103001(2012),Phys.Rev.Lett.123:140401(2019),PRX QUANTUM 3.030342 (2022),PhysRevA.106.042206(2022)}, ring-network configuration \cite{Quantum Inf.Process.16.266(2017)}, tree-network configuration \cite{PhysRevA.104.042405(2021)}, and so on.
More recently, Yang \emph{et al} \cite{Entropy.24.691(2022)} proposed an $n$-layer tree-network, which involves odd-numbered (centrosymmetric) chain-network, star-network and any positive-integer-forked tree-network. However, such layered arrangement may be inherently difficult to describe centerless networks (an even-numbered chain-network, for example).
In fact, although there exist various quantum networks, finding a universal network architecture is still an open problem.

In this paper, we propose an $(n,m,p)$-type quantum network configuration and derive the corresponding $n$-local inequality based on the assumption of independent sources.
It can cover most of the existing acyclic networks, for example, the typical chain-network and star-network configurations.
Then we show the non-$n$-locality of our network by calculating the violation of the $n$-local inequality with bipartite entangled sources and Pauli measurements on these distant network nodes.

\section{The bilocal network with two independent sources}

\begin{figure}[h]
      \centering
      \includegraphics[width=8cm, height=3.5cm]{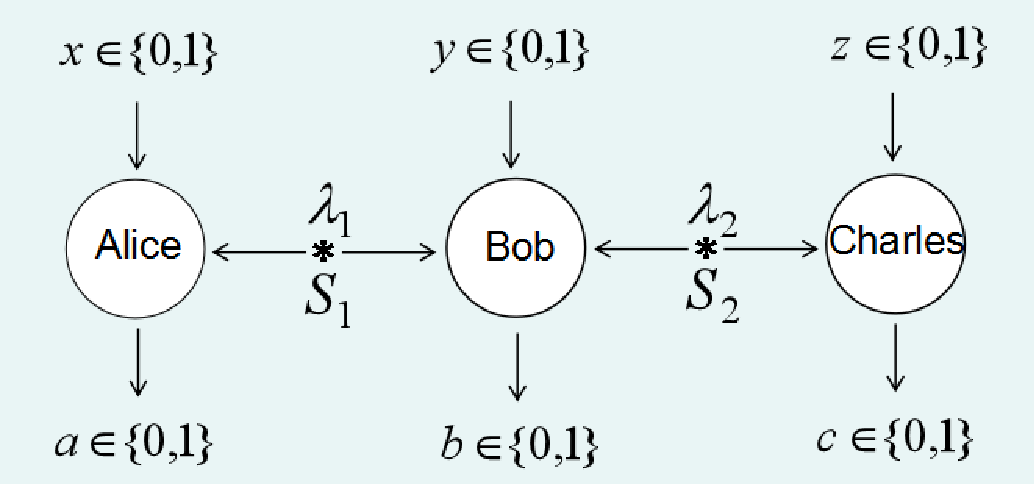}
      \caption{The bilocal network.}
\label{bilocal-network}
\end{figure}

In this section, we describe the bilocal network with two independent sources \cite{Phys.Rev.Lett.104:170401(2010),PhysRevA.85.032119(2012)}, as shown in Fig.\ref{bilocal-network}.
Suppose that there are two independent (space-like separated) two-particle sources $S_{1}$ and $S_{2}$ related to two classical variables $\lambda_{1}$ and $\lambda_{2}$, respectively, with a joint distribution $\rho(\lambda_{1},\lambda_{2})=\rho_{1}(\lambda_{1})\rho_{2}(\lambda_{2})$ satisfying $\int \rho_{1}d\lambda_{1}=\int \rho_{2}d\lambda_{2}=1$.
The source $S_{1}$ sends particles to Alice and Bob, and the source $S_{2}$ sends particles to Bob and Charles. We can call Bob and Alice (or Charles) the central party (node) and extremal party, respectively. So there are one central party and two extremal parties in this bilocal network.
The parties Alice, Bob and Charles perform the dichotomic measurements with inputs (outputs) $x$, $y$ and $z$ ($a$, $b$ and $c$), respectively.

A correlation between the measurement outcomes is usually characterized by the joint probability distribution.
In the bilocal network, the joint probability distribution can be described by
\begin{eqnarray}\label{P1}
P(a,b,c|x,y,z)=\int\int d\lambda_{1}d\lambda_{2} \rho_{1}(\lambda_{1})\rho_{2}(\lambda_{2})P(a|x,\lambda_{1})P(b|y,\lambda_{1},\lambda_{2})P(c|z,\lambda_{2}),
\end{eqnarray}
where $P(a,b,c|x,y,z)$ is the joint probability distribution for the outcomes from all of the three parties and $P(a|x,\lambda_{1}), P(b|y,\lambda_{1},\lambda_{2})$ and $P(c|z,\lambda_{2})$ are respectively for Alice, Bob and Charles.

Let $I=\frac{1}{4}\sum_{x,z}\langle A_{x}B_{0}C_{z} \rangle$, $J=\frac{1}{4}\sum_{x,z}(-1)^{x+z}\langle A_{x}B_{1}C_{z} \rangle$, and $\langle A_{x}B_{y}C_{z} \rangle=\sum_{a,b,c}(-1)^{a+b+c}P(a,b,c|x,y,z)$, where $ A_{x}$, $B_{y}$ and $C_{z}$ are the observables related to parties Alice, Bob and Charles, respectively. By this, one can obtain the bilocal inequalities \cite{PhysRevA.85.032119(2012)}
\begin{eqnarray}\label{}
\sqrt{I}+\sqrt{J}\leq1.
\end{eqnarray}
Different from the Bell inequalities, these are nonlinear inequalities.
The quantum network is bilocal correlation if these inequalities hold for any quantum systems, otherwise non-bilocal correlation.

\section{$(n,m,p)$-type network configuration}

\subsection{The network configuration}

In quantum network, we call all of the parties \textit{network nodes} and they are divided into two categories: the extremal nodes and the intermediate (non-extremal) nodes.
We next present an ($n,m,p$)-type acyclic network configuration, where $n$ is the number of the independent sources, $m$ represents the number of particles owned by each intermediate node, and $p$ is the number of the extremal nodes.

We here restrict our attention to two-particle sources and dichotomic measurements for clarity.
Since each source $S_{r}$, characterized by a classical variable $\lambda_{r}$, $r=1,2,\cdots,n$, can only shared between two neighboring nodes, each intermediate node will be related to $m$ distant sources and each extremal node connects one source.
Each source provides two particles, so there are $2n$ particles, $l=(2n-p)/{m}$ intermediate nodes and $p\leq n$ extremal nodes in our network configuration.

\begin{figure}[h]
      \centering
      \subfigure[ ~chain-network ($n,2,2$)]
           {
           \label{}
           \includegraphics[width=8.2cm, height=4cm]{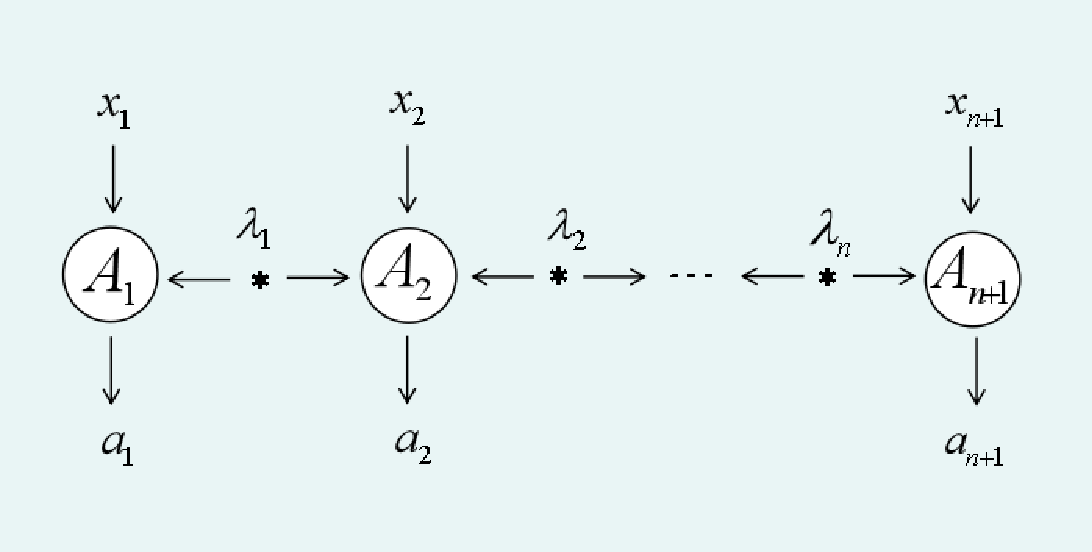}
           }
     \hspace{0.0005in}
      \subfigure[ ~star-network ($n,n,n$)]
           {
           \label{}
           \includegraphics[width=6.2cm, height=4cm]{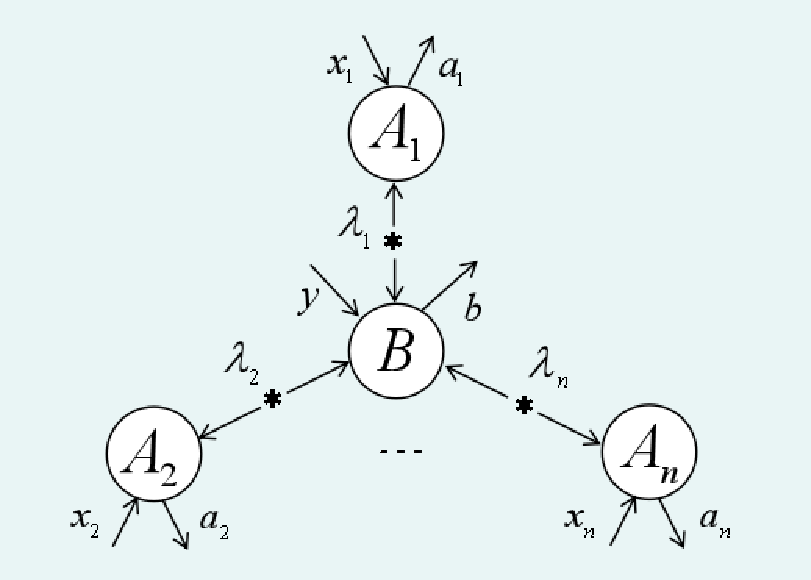}
           }
           \hspace{0.0005in}
      \subfigure[ ~tree-network ($n,m,n-(n-m)/(m-1)$)]
           {
           \label{}
           \includegraphics[width=8.2cm, height=5.8cm]{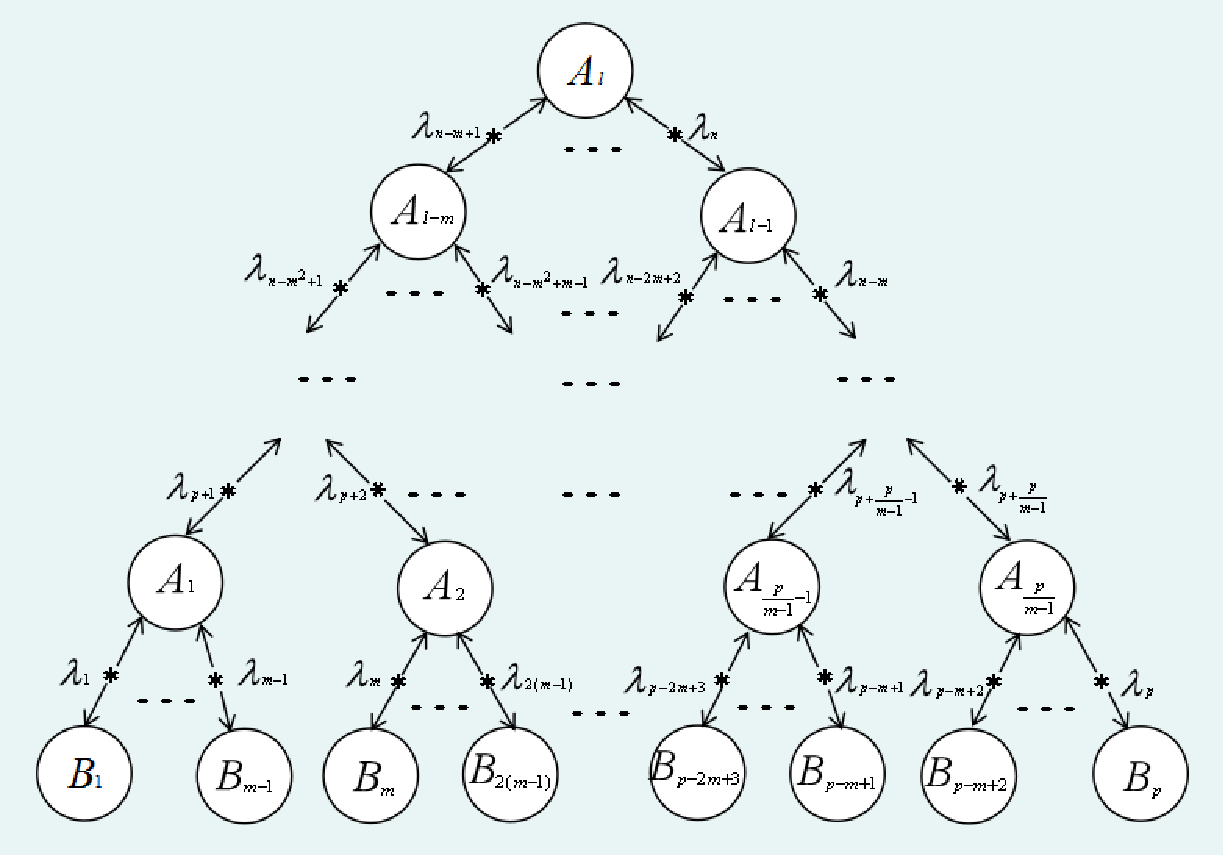}
           }
           \hspace{0.0005in}
      \subfigure[ ~centerless and asymmetric network ($15,3,9$)]
           {
           \label{}
           \includegraphics[width=6.2cm, height=5.8cm]{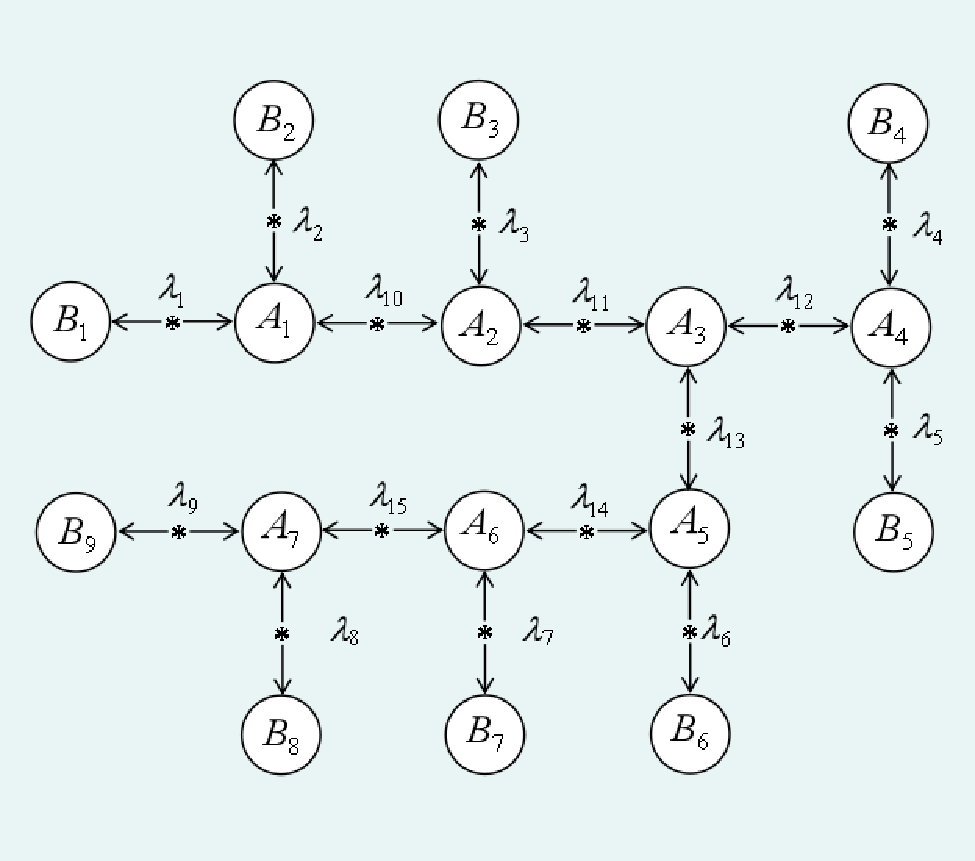}
           }
      \caption{The $n$-local scenarios for (a) chain-network ($n,2,2$), (b) star-network ($n,n,n$), (c) layered tree-network ($n,m,n-(n-m)/(m-1)$), and (d) an arbitrary centerless and asymmetric network ($15,3,9$).}
\label{n-local-network}
\end{figure}

Intuitively, we provide the typical (a) chain-network ($n,2,2$), (b) star-network ($n,n,n$), (c) layered tree-network ($n,m,n-(n-m)/(m-1)$), and (d) an arbitrary centerless and asymmetric network ($15,3,9$) for example, as shown in Fig.\ref{n-local-network}.

\subsection{$n$-local correlation inequalities}

For $n\geq 2$, let $A^{i}(i=1,2,\cdots,l)$ and $B^{j}(j=1,2,\cdots, p)$ are respectively the dichotomic measurement operators of the intermediate nodes and extremal nodes.
$x_{i}$ and $a_{i}$ are respectively the inputs and outputs for $A^{i}$, and $y_{j}$ and $b_{j}$ for $B^{j}$, with $x_{i},y_{j},a_{i},b_{j}\in\{0,1\}$.
Let $X=\{x_{1},x_{2},\cdots,x_{l}\}$, $Y=\{y_{1},y_{2},\cdots,y_{p}\}$, $A=\{a_{1},a_{2},\cdots,a_{l}\}$ and $B=\{b_{1},b_{2},\cdots,b_{p}\}$ be sets of the inputs and outputs, for simplicity.

The independent source assumption enables us to write the joint distribution of the classical variables $\lambda_{1},\lambda_{2},\cdots,\lambda_{n}$ as $\rho(\lambda_{1})\rho(\lambda_{2})\cdots\rho(\lambda_{n})$ satisfying $\int \rho(\lambda_{r})d\lambda_{r}=1, r=1,2,\cdots,n$.
Let $P(A,B|X,Y)$ is the joint probability distribution for the outcomes of all nodes, and $P(a_{i}|x_{i},\Lambda_{i})$ and $P(b_{j}|y_{j},\lambda_{j})$ are respectively for nodes $A^{i}(i=1,2,\cdots,l)$ and $B^{j}(j=1,2,\cdots, p)$, where $\Lambda_{i}$ denotes the set of classical variables that arrive to node $A^{i}$.
If the joint probability distribution can be expressed as
\begin{eqnarray}\label{}
 P(A,B|X,Y)
&=&\int\cdots\int d\lambda_{1}d\lambda_{2}\cdots d\lambda_{n}\rho(\lambda_{1})\rho(\lambda_{2})\cdots\rho(\lambda_{n})
  P(a_{1}|x_{1},\Lambda_{1}) \times \cdots \times  P(a_{i}|x_{i},\Lambda_{i}) \nonumber\\ &&
\times \cdots \times
 P(a_{l}|x_{l},\Lambda_{l})P(b_{1}|y_{1},\lambda_{1})P(b_{2}|y_{2},\lambda_{2}) \times \cdots \times P(b_{p}|y_{p},\lambda_{p}),
\end{eqnarray}
then this network is $n$-local, otherwise it is non-$n$-local.
Taking layered tree-network ($n,m,n-(n-m)/(m-1)$) for example, see Fig.\ref{n-local-network} (c). One may take
\begin{eqnarray}\label{}
 P(A,B|X,Y)
&=&\int\cdots\int d\lambda_{1}d\lambda_{2}\cdots d\lambda_{n}\rho(\lambda_{1})\rho(\lambda_{2})\cdots\rho(\lambda_{n})
  P(a_{1}|x_{1},\lambda_{1},\cdots,\lambda_{m-1},\lambda_{p+1}) \nonumber\\ &&
 \times P(a_{2}|x_{2},\lambda_{m},\cdots,\lambda_{2(m-1)},\lambda_{p+2})
 \times \cdots \times P(a_{t}|x_{t},\lambda_{(t-1)(m-1)+1},\cdots,\lambda_{t(m-1)},\lambda_{p+t}) \nonumber\\ &&
 \times \cdots \times P(a_{l}|x_{l},\lambda_{n-m+1},\cdots,\lambda_{n})P(b_{1}|y_{1},\lambda_{1})P(b_{2}|y_{2},\lambda_{2})\cdots P(b_{p}|y_{p},\lambda_{p}),
\end{eqnarray}
where $t\in \{0,1,\cdots,n-p\}$ (the index of classical variable $\lambda_{p+t}$) satisfy $p+t\leq n$.

To proceed, let $k=0,1$, and define
\begin{eqnarray}\label{}
I^{k}_{X}=\frac{1}{2^{p}}\sum_{Y} (-1)^{\sum_{j=1}^{p}ky_{j}}\langle A_{x_{1}}^{1}A_{x_{2}}^{2}\cdots A_{x_{l}}^{l}B_{y_{1}}^{1}B_{y_{2}}^{2}\cdots B_{y_{p}}^{p} \rangle,
\end{eqnarray}
where the correlator
\begin{eqnarray}\label{}
\langle A_{x_{1}}^{1}A_{x_{2}}^{2}\cdots A_{x_{l}}^{l}B_{y_{1}}^{1}B_{y_{2}}^{2}\cdots  B_{y_{p}}^{p} \rangle=\sum_{A,B} (-1)^{\sum_{i=1}^{l}a_{i}+\sum_{j=1}^{p}b_{j}}  P(A,B|X,Y).
\end{eqnarray}

Then we calculate the absolute value of this linear combination $|I^{k}_{X}|$ and have
\begin{eqnarray}\label{}
|I^{k}_{X}| &=& \frac{1}{2^{p}}|\sum_{Y} (-1)^{\sum_{j=1}^{p}ky_{j}}\langle A_{x_{1}}^{1}A_{x_{2}}^{2}\cdots A_{x_{l}}^{l}B_{y_{1}}^{1}B_{y_{2}}^{2}\cdots B_{y_{p}}^{p} \rangle| \nonumber\\
&=&\frac{1}{2^{p}}|\sum_{Y} (-1)^{\sum_{j=1}^{p}ky_{j}}
\int\cdots\int d\lambda_{1}d\lambda_{2}\cdots d\lambda_{n}
\rho(\lambda_{1})\rho(\lambda_{2})\cdots\rho(\lambda_{n})
\sum_{a_{1}}(-1)^{a_{1}}P(a_{1}|x_{1},\Lambda_{1})\nonumber\\ & &
\times\cdots \times \sum_{a_{i}}(-1)^{a_{i}}P(a_{i}|x_{i},\Lambda_{i})\times\cdots
\times \sum_{a_{l}}(-1)^{a_{l}}P(a_{l}|x_{l},\Lambda_{l})\times\sum_{b_{1}}(-1)^{b_{1}}P(b_{1}|y_{1},\lambda_{1})\nonumber\\ & &
\times\cdots \times \sum_{b_{j}}(-1)^{b_{j}}P(b_{j}|y_{j},\lambda_{j})\times\cdots
\times \sum_{b_{p}}(-1)^{b_{p}}P(b_{p}|y_{p},\lambda_{p})|\nonumber\\
&=&\frac{1}{2^{p}}|\sum_{Y} (-1)^{\sum_{j=1}^{p}ky_{j}}
\int\cdots\int d\lambda_{1}d\lambda_{2}\cdots d\lambda_{n}\rho(\lambda_{1})\rho(\lambda_{2})\cdots\rho(\lambda_{n})
\langle A_{x_{1}}^{1}\rangle\langle A_{x_{2}}^{2}\rangle
\cdots\langle A_{x_{l}}^{l}\rangle\langle B_{y_{1}}^{1}\rangle\langle B_{y_{2}}^{2}\rangle\cdots\langle B_{y_{p}}^{p}\rangle|\nonumber\\
& \leq &\frac{1}{2^{p}} \int\cdots\int d\lambda_{1}d\lambda_{2}\cdots d\lambda_{n}\rho(\lambda_{1})\rho(\lambda_{2})\cdots\rho(\lambda_{n})
\prod_{i=1}^{l} |\langle A_{x_{i}}^{i}\rangle| \prod_{j=1}^{p}|\langle B_{0}^{j}\rangle+(-1)^{k}\langle B_{1}^{j}\rangle|.
\end{eqnarray}
Noting that $|\langle A_{x_{i}}^{i}\rangle|\leq1$ we have
\begin{eqnarray}\label{}
|I^{k}_{X}|& \leq &\prod_{j=1}^{p}\int d\lambda_{j}\rho(\lambda_{j})\frac{1}{2}|\langle B_{0}^{j}\rangle+(-1)^{k}\langle B_{1}^{j}\rangle|.
\end{eqnarray}
As a result, it seems likely that the upper bound of the linear combination $|I^{k}_{X}|$ is only depended on the extremal nodes, no matter what choice of $A_{x_{i}}^{i}$ we make for each intermediate node $A_{i}^{}$.

Using the Cauchy inequality \cite{PhysRevA.85.032119(2012)}
\begin{eqnarray}\label{Cauchy}
\sum_{k=1} ^{m}(\prod_{i=1} ^{n} x_{i} ^{k})^\frac{1}{n}\leq\prod_{i=1} ^{n}(x_{i} ^{1}+x_{i} ^{2}+\ldots +x_{i} ^{m})^{\frac{1}{n}},
\end{eqnarray}
one can obtain
\begin{eqnarray}\label{}
|I_{X}^{0}|^{\frac{1}{p}}+|I_{X'}^{1}|^{\frac{1}{p}}&\leq &\prod_{j=1}^{p}[\int d\lambda_{j}\rho(\lambda_{j})\frac{1}{2}(|\langle B_{0}^{j}\rangle+\langle B_{1}^{j}\rangle|+|\langle B_{0}^{j}\rangle-\langle B_{1}^{j}\rangle|)]^{\frac{1}{p}}\nonumber\\
&=&\prod_{j=1}^{p}[\int d\lambda_{j}\rho(\lambda_{j}) \text{max}\{|\langle B_{0}^{j}\rangle|,|\langle B_{1}^{j}\rangle|\}]^{\frac{1}{p}},
\end{eqnarray}
where $X, X' =\{x_{1},x_{2},\cdots,x_{l}\}$ are respectively the sets of the binary inputs for the intermediate nodes.

Since $|\langle B_{0}^{j}\rangle|\leq1$ and $|\langle B_{1}^{j}\rangle|\leq1$, we have
\begin{eqnarray}\label{}
|I_{X}^{0}|^{\frac{1}{p}}+|I_{X'}^{1}|^{\frac{1}{p}}\leq 1.
\end{eqnarray}
This is a set of $n$-local nonlinear correlation inequalities derived from the assumption of independent sources.
Any violation of these inequalities by physical system will be seen as a nonlocality witness for the present network.

\section{Quantum violations of the $n$-local inequalities}

Let us now construct explicitly the quantum distribution and verify the nonlocality of the present $(n,m,p)$-type quantum networks.
Suppose that each quantum source produces two-qubit pure state $|\Psi_{r}\rangle=\cos\theta_{r}|00\rangle+\sin\theta_{r}|11\rangle$, $\theta_{r}\in[0,2\pi]$, $r=1,2,\cdots,n$.
Let $\rho_{r}=|\Psi_{r}\rangle \langle\Psi_{r}|$, then this ($2n$-qubit) complex system will be $\rho=\bigotimes_{r=1}^{n}\rho_{r}$.
Note that each intermediate node $A^{i}$ receives $m$ independent qubits and each extremal node $B^{j}$ only one qubit.

Consider the dichotomic measurement, where each intermediate node performs the joint measurements $A_{0}=\sigma_{z}^{\otimes m}$ or $A_{1}=\sigma_{x}^{\otimes m}$ on its local qubits and each extremal node performs a local measurement $B_{y_{j}}^{j}=\cos\alpha_{j}\sigma_{z}+(-1)^{y_{j}}\sin \alpha_{j}\sigma_{x}$, $\alpha_{j}\in[0,2\pi]$, $j=1,2,\cdots,p$, on its single-qubit subsystem, where $\sigma_{x}$ and $\sigma_{z}$ are the Pauli operators in the computational basis.

We here calculate $S=|I^{0}_{0,0,\cdots,0}|^{{1}/{p}}+|I^{1}_{1,1,\cdots,1}|^{{1}/{p}}$ for clarity.
Then we have
\begin{eqnarray}\label{}
I^{0}_{0,0,\cdots,0}&=&\frac{1}{2^{p}}\sum_{Y} \langle A_{0}^{1}A_{0}^{2}\cdots A_{0}^{l}B_{y_{1}}^{1}B_{y_{2}}^{2}\cdots B_{y_{p}}^{p} \rangle\nonumber\\
&=&\text{Tr}[ \rho \sigma_{z}^{\otimes (2n-p)} \prod_{j=1}^{p}\cos\alpha_{j} \sigma_{z}^{\otimes p}]\nonumber\\
&=&\prod_{j=1}^{p}\cos\alpha_{j}
\end{eqnarray}
and
\begin{eqnarray}\label{}
I^{1}_{1,1,\cdots,1}&=&\frac{1}{2^{p}}\sum_{Y} (-1)^{\sum_{j=1}^{p} y_{j}}\langle A_{1}^{1}A_{1}^{2}\cdots A_{1}^{l}B_{y_{1}}^{1}B_{y_{2}}^{2}\cdots B_{y_{p}}^{p} \rangle\nonumber\\
&=&\text{Tr}[ \rho \sigma_{x}^{\otimes (2n-p)} \prod_{j=1}^{p}\sin\alpha_{j} \sigma_{x}^{\otimes p}]\nonumber\\
&=&\prod_{j=1}^{p}\sin\alpha_{j}\prod_{r=1}^{n}\sin(2\theta_{r}).
\end{eqnarray}
So we obtain
\begin{eqnarray}\label{S}
S=|\prod_{j=1}^{p}(\cos\alpha_{j})|^{\frac{1}{p}}+|\prod_{j=1}^{p}\sin\alpha_{j}\prod_{r=1}^{n}\sin(2\theta_{r})|^{\frac{1}{p}}.
\end{eqnarray}

Without loss of generality, let $\alpha_{j}=\alpha$, then one gets
\begin{eqnarray}
S=|\cos\alpha|+|\sin\alpha| |\prod_{r=1}^{n}\sin(2\theta_{r})|^{\frac{1}{p}}.
\end{eqnarray}
Let $\partial{S}/\partial{\alpha}=0$, and thus we have $|\prod_{r=1}^{n}\sin(2\theta_{r})|^{{1}/{p}}=|\tan\alpha|$.
Therefore, we can obtain
\begin{eqnarray}
S_{\text{max}}=\sqrt{1+[\prod_{r=1}^{n}\sin^{2}(2\theta_{r})]^{\frac{1}{p}}} \geq 1.
\end{eqnarray}
The maximum value of $S_{\text{max}}$ is $\sqrt{2}$, and it occurs at $\theta_{r}=\pi/4$ or an odd multiple of $\pi/4$; the minimum value 1 occurs at $\theta_{r}=0$ or integer multiple of $\pi/2$.
It means that our $(n,m,p)$-type quantum network is non-$n$-local if and only if the given all bipartite quantum sources are entangled, i.e., the concurrence \cite{Rev.Mod.Phys.81:865(2009)} $C(|\Psi_{r}\rangle)=|\sin(2\theta_{r})|>0$.

\section{Discussion and summary}

In summary, based on the assumption of independent source we have constructed an $(n,m,p)$-type network configuration. A set of nonlinear $n$-local correlation inequalities have been derived. Given a set of bipartite entangled sources we calculate the quantum prediction with the dichotomic measurements consisting of the Pauli operators. Consequently, the $n$-local inequality will be obviously violated conditional on these quantum sources are entangled. The non-$n$-locality of the present $(n,m,p)$-type quantum network have been demonstrated.

Note that there are two main results in our work.
First, our $(n,m,p)$-type network is a universal network configuration. It can provide a practical tool to investigate the non-$n$-locality of an arbitrary $(n,m,p)$-type network. For example, the seminal bilocal scenario \cite{Phys.Rev.Lett.104:170401(2010)} corresponds to $(2,2,2)$;  the chain-network involving arbitrary nodes \cite{Quantum Inf.Process.14.2025(2015)} corresponds to $(n,2,2)$; the star-network configuration \cite{PhysRevA.90.062109(2014)} corresponds to $(n,n,n)$; the recent layered tree-network configuration \cite{Entropy.24.691(2022)} given the specified $m$ as ($n,m,n-(n-m)/(m-1)$), and so on.
Second, since there is no special central node in the present configuration, all intermediate nodes are equivalent and thus our quantum network can meet both centerless and asymmetric configuration. Its expansibility follows from the fact that these $n$-local inequalities only depend on the number of independent sources and extremal nodes.
Meanwhile, it does not involve the cases that the number $m$ is allowed to be different or varied. Although the number $m$ does not emerge in nonlinear inequalities, it does really affect the network designing.
We expect that this work will motivate further investigating on the quantum network configurations and provide a way to construct more universal quantum networks for quantum information processing.

\begin{acknowledgements}
This work was supported by
the National Natural Science Foundation of China under Grant Nos: 62271189, 12071110,
the Hebei Central Guidance on Local Science and Technology Development Foundation of China under Grant Nos: 226Z0901G, 236Z7604G,
the Hebei 3-3-3 Fostering Talents Foundation of China under Grant No: A202101002,
the Education Department of Hebei Province Natural Science Foundation of China under Grant No: ZD2021407,
the Education Department of Hebei Province Teaching Research Foundation of China under Grant No: 2021GJJG482.
\end{acknowledgements}


\begin{thebibliography}{}
\bibitem{Physics.1(3):195-200(1964)} J. S. Bell, On the Einstein-Podolsky-Rosen paradox, Physics {\bf 1}, 195 (1964).
\bibitem{Rev.Mod.Phys.82.665(2010)} H. Buhrman, R. Cleve, S. Massar, and R. de Wolf, Nonlocality and communication complexity, Rev. Mod. Phys. {\bf 82}, 665 (2010).
\bibitem{Rev.Mod.Phys.86:419(2014)} N. Brunner, D. Cavalcanti, S. Pironio, V. Scarani, and S. Wehner, Bell nonlocality, Rev. Mod. Phys. {\bf 86}, 419 (2014).
\bibitem{Rev.Mod.Phys.81:865(2009)} R. Horodecki, P. Horodecki, M. Horodecki, and K. Horodecki,  Quantum entanglement, Rev. Mod. Phys. {\bf 81}, 865 (2009).
\bibitem{NC2000} M. A. Nielsen and I. L. Chuang, {\it Quantum Computation and Quantum Information} (Cambridge University Press, Cambridge, 2000).
\bibitem{Phys.Rev.Lett.87.117901(2001)} V. Scarani and N. Gisin, Quantum communication between $n$ partners and Bell's inequalities, Phys. Rev. Lett. {\bf87}, 117901 (2001).

\bibitem{Phys.Rev.Lett.97.120405(2006)}  A. Ac\'{\i}n, N. Gisin, and L. Masanes, From Bell's theorem to secure quantum key distribution, Phys. Rev. Lett. {\bf 97}, 120405 (2006).
\bibitem{M2015} M. Hayashi, S. Ishizaka, A. Kawachi, G. Kimura, and T. Ogawa, {\it Introduction to Quantum Information Science} (Springer, Berlin Heidelberg, 2015).
\bibitem{GYE-PRL2014} T. Gao, F. L. Yan, and S. J. van Enk, Permutationally invariant part of a density matrix and nonseparability of $n$-qubit states, Phys. Rev. Lett. {\bf 112}, 180501 (2014).
\bibitem{DHYG-JPA2020} D. Ding, Y. Q. He, F. L. Yan, and T. Gao, Optimizing dichotomic local phase measurement settings for multipartite quantum systems, J. Phys. A-Math. Theor. {\bf 53}, 265301 (2020).
\bibitem{Phys.Rev.Lett.104:170401(2010)} C. Branciard, N. Gisin, and S. Pironio, Characterizing the nonlocal correlations created via entanglement swapping, Phys. Rev. Lett. {\bf 104}, 170401 (2010).


\bibitem{Phys.Rev.Lett.71.4287(1993)} M. \.{Z}ukowski, A. Zeilinger, M. A. Horne, and A. K. Ekert, ``Event-ready-detectors" Bell experiment via entanglement swapping, Phys. Rev. Lett. {\bf 71}, 4287 (1993).
\bibitem{PhysRevA.85.032119(2012)} C. Branciard, D. Rosset, N. Gisin, and S. Pironio, Bilocal versus nonbilocal correlations in entanglement-swapping experiments, Phys. Rev. A {\bf 85}, 032119 (2012).
\bibitem{PhysRevLett.120.140402(2018)} M. X. Luo, Computationally efficient nonlinear Bell inequalities for quantum networks, Phys. Rev. Lett. {\bf 120}, 140402 (2018).

\bibitem{Tavakoli-network2022} A. Tavakoli, A. Pozas-Kerstjens, M. X. Luo, and M. O. Renou, Bell nonlocality in networks, Rep. Prog. Phys. {\bf 85}, 056001 (2022).
\bibitem{PhysRevLett.128.010403(2022)} A. Pozas-Kerstjens, N. Gisin, and A. Tavakoli, Full network nonlocality, Phys. Rev. Lett. {\bf 128}, 010403 (2022).

\bibitem{Phys.Rev.A.105:042436.(2022)} W. L. Hou, X. W. Liu, and C. L. Ren, Network nonlocality sharing via weak measurements in the extended bilocal scenario, Phys. Rev. A {\bf 105}, 042436 (2022).
\bibitem{Quantum Inf.Process.14.2025(2015)} K. Mukherjee, B. Paul, and D. Sarkar, Correlations in $n$-local scenario, Quantum Inf. Process. {\bf 14}, 2025 (2015).

\bibitem{PhysRevA.102.052222} A. Kundu, M. K. Molla, I. Chattopadhyay, and D. Sarkar, Maximal qubit violation of $n$-local inequalities in a quantum network, Phys. Rev. A {\bf 102}, 052222 (2020).
\bibitem{PhysRevA.107.032404 (2023)} K. Mukherjee, I. Chakrabarty, and G. Mylavarapu, Persistency of non-$n$-local correlations in noisy linear networks, Phys. Rev. A {\bf 107}, 032404 (2023).
\bibitem{PhysRevA.108.032416 (2023)} K. Mukherjee, S. Mandal, T. Patro, and N. Ganguly, Hidden non-$n$-locality in linear networks, Phys. Rev. A {\bf 108}, 032416 (2023).

\bibitem{PhysRevA.90.062109(2014)} A. Tavakoli, P. Skrzypczyk, D. Cavalcanti, and A. Ac\'{\i}n, Nonlocal correlations in the star-network configuration, Phys. Rev. A {\bf 90}, 062109 (2014).
\bibitem{New J.Phys.19.073003(2017)} A. Tavakoli, M. O. Renou, N. Gisin, and N. Brunner, Correlations in star networks: from Bell inequalities to network inequalities, New J. Phys. {\bf 19}, 073003 (2017).
\bibitem{New J.Phys.19.113020.(2017)} F. Andreoli, G. Carvacho, L. Santodonato, R. Chaves, and F. Sciarrino, Maximal qubit violation of $n$-locality inequalities in a star-shaped quantum network, New J. Phys. {\bf 19}, 113020 (2017).
\bibitem{PhysRevA.108.042409 (2023)} B. Doolittle and E. Chitambar, Maximal qubit violations of $n$-locality in star and chain networks, Phys. Rev. A {\bf 108}, 042409 (2023).
\bibitem{New J.Phys.14.103001(2012)} T. Fritz, Beyond Bell's theorem: correlation scenarios, New J. Phys. {\bf 14}, 103001 (2012).
\bibitem{Phys.Rev.Lett.123:140401(2019)} M. O. Renou, E. B\"{a}umer, S. Boreiri, N. Brunner, N. Gisin, and S. Beigi, Genuine quantum nonlocality in the triangle network, Phys. Rev. Lett. {\bf 123}, 140401 (2019).
\bibitem{PRX QUANTUM 3.030342 (2022)} A. Suprano, D. Poderini, E. Polino, I. Agresti, G. Carvacho, A. Canabarro, E. Wolfe, R. Chaves, and F. Sciarrino, Experimental genuine tripartite nonlocality in a quantum triangle network, Phys. Rev. X {\bf 3}, 030342 (2022).
\bibitem{PhysRevA.106.042206(2022)} K. Mukherjee, Detecting nontrilocal correlations in a triangle network, Phys. Rev. A {\bf 106}, 042206 (2022).

\bibitem{Quantum Inf.Process.16.266(2017)} M. Frey, A Bell inequality for a class of multilocal ring networks, Quantum Inf. Process. {\bf16}, 266 (2017).
\bibitem{PhysRevA.104.042405(2021)} L. H. Yang, X. F. Qi, and J. C. Hou, Nonlocal correlations in the tree-tensor-network configuration, Phys. Rev. A {\bf 104}, 042405 (2021).
\bibitem{Entropy.24.691(2022)} L. H. Yang, X. F. Qi, and J. C. Hou, Quantum nonlocality in any forked tree-shaped network, Entropy {\bf24}, 691 (2022).

\end{thebibliography}
\end{document}